\documentclass{ws-ijmpd}

\begin{document}

\markboth{L. Arturo Ure\~na-L\'opez and Mayra J. Reyes-Ibarra}
{On the dynamics of a quadratic scalar field potential}

%
\catchline{}{}{}{}{}
%

\title{ON THE DYNAMICS OF A QUADRATIC SCALAR FIELD POTENTIAL}

\author{L. Arturo Ure\~na-L\'opez and Mayra J. Reyes-Ibarra}

\address{Departamento de F\'isica, DCI, Campus Le\'on, Universidad de
  Guanajuato, \\
C.P. 37150, Le\'on, Guanjuato, M\'exico\\
lurena@fisica.ugto.mx\\
mreyes@fisica.ugto.mx}

\maketitle


\begin{abstract}
We review the attractor properties of the simplest chaotic model of
inflation, in which a minimally coupled scalar field is endowed with a
quadratic scalar potential. The equations of motion in a flat
Friedmann-Robertson-Walker universe are written as an autonomous
system of equations, and the solutions of physical interest
appear as critical points. This new formalism is then applied to the
study of inflation dynamics, in which we can go beyond the known
slow-roll approximation.
\end{abstract}


\section{Introduction} \label{sec:introduction}	
One of the most studied issues in Cosmology is the dynamics of
cosmological scalar fields, mostly because of their usefulness in
providing models for different needed processes in the evolution of
the universe. It is not an easy task at all to present a full
bibliography on scalar fields, but a nice and comprehensive review was
recently presented by Copeland, Sami and Tsujikawa in
Ref~\refcite{Copeland:2006wr}.

A full plethora of methods exist in the specialized literature to
study the dynamics of cosmological scalar fields, but only a few can
be told to be extreme useful and of general applicability, in a field
where exact solutions are rarely found.

One of these methods is the writing of the scalar and gravitational
evolution equations in the form of a \emph{dynamical system}. This is
an idea that has largely pervaded the specialized literature on cosmological
scalar fields\cite{Belinsky:1985zd,Copeland:1997et,Coley:1999uh}. The
reason is that the theory of dynamical systems can show the existence
of (fixed) stationary points that may represent important cosmological
solutions; many times these solutions are attractors the system evolves to
independently of the initial conditions. This is of wide interest in
the case of inflationary and dark energy
models\cite{Copeland:2006wr}. However, the dynamical system approach
is not always completely suitable for the study of scalar field
dynamics.

A typical example is a scalar field model endowed with a quadratic
potential. The first attempt for this case was
made by Belinsky et al\cite{Belinsky:1985zd} (see
also\cite{Starobinsky78}), in which the dynamical variables were just
the normalized values of the scalar field  and gravitational variables
(the scalar field value $\phi$, its time derivative $\dot{\phi}$, and
the Hubble parameter $H$). Their analysis did not reveal inflationary
attractor solutions, but rather asymptotic inflationary behavior at
times $t \to -\infty$, for which the corresponding fixed points were
all unstable.

In a later paper, de la Macorra and
Piccinelli\cite{delaMacorra:1999ff} used the same variables suggested
in Ref.~\refcite{Copeland:1997et} for an exponential potential; for the
latter is possible to write the equations of motion as an autonomous
plane system. On the contrary, the system of equations for a quadratic
potential case cannot be written in such a simple form, and one has to
introduce extra variables, which are not directly related with the
original dynamical variables, to have a closed system of equations. In
any case, de la Macorra and Piccinelli were able to obtain asymptotic
solutions, and provided a general classification for the behavior of
arbitrary scalar potentials\cite{delaMacorra:1999ff}.

Our own proposal to study the dynamics of a quadratic potential is
somehow in between the last two discussed above. We take the same
variables used in\cite{Copeland:1997et,delaMacorra:1999ff}, and define
a third new variable related to the Hubble parameter; the resulting dynamical
system is a 3-dimensional autonomous one. However, the nature of the
fixed points of physical interest is clearly revealed if one studies a
reduced 2-dimensional dynamical system, and the third variable is
just taken as a \emph{control} parameter. 

The critical points that will be considered are not fixed in the
strict sense, as their location on the 2-dimensional phase space
depend upon the values of the control variable. Nevertheless, we shall
show that the critical points of the reduced dynamical system carry
useful physical information, and that their stability can be
established by standard methods.

A similar approach was presented in a recent paper
in\cite{Chongchitnan:2007eb}, in which the dynamics of quintessence
models is under the control of a so-called \emph{roll} parameter
$\lambda \propto V^\prime/V$. The location of the fixed points, which
are actually those found for an exponential
potential\cite{Copeland:1997et}, is determined by the value of
$\lambda$ at any instant of time. Such a property provides a powerful
method to study general quintessence models.

Even though a quadratic potential is the simplest and the most well known
 of the inflationary
 models\cite{Linde:1983gd,Liddle:2000cg,Bassett:2005xm}, we will
 revise some inflationary solutions from the point of view of the
 results on attractor solutions obtained from the dynamical system
 approach.

A summary of the paper is as follows. In Sec.~\ref{sec:math-backgr},
we describe the mathematical definitions of the new dynamical system,
and make a detailed study of its critical points. In
Sec.~\ref{sec:infl-dynam}, we apply our results to inflationary models
in the simplest chaotic scenario. Finally, conclusions are discussed
in Sec.~\ref{sec:conclusions}.

\section{Mathematical background}\label{sec:math-backgr}
The model is defined by the following action in the scalar field $\phi$,
\begin{equation}
  S_\phi =- \int dx^{4}\sqrt{-g}\left[\frac{1}{2} \partial^{\mu} \phi
  \partial_{\mu} \phi + \frac{1}{2} m^2 \phi^2 \right] \, ,
\end{equation}
where $m$ indicates the scalar field mass. The evolution equations in
a spatially flat FRW model, with a metric given by $g_{\mu \nu} =
\mathrm{diag}(-1,a^2(t),a^2(t),a^2(t))$, are
\begin{subequations}
\label{eq:eqs-motion}
\begin{eqnarray}
  \ddot{\phi} &=& -3H\dot{\phi} - m^2 \phi \, ,
  \label{eq:eqs-motion-a} \\
  \dot{H} &=& - 4\pi G \dot{\phi}^2 \, , \label{eq:eqs-motion-b}
\end{eqnarray}
\end{subequations}
together with the Friedmann constraint
\begin{equation}
  H^2\equiv \left( \frac{\dot{a}}{a} \right)^2 = \frac{4\pi G }{3}
  \left( \dot{\phi}^2 + m^2 \phi^2 \right) \, . \label{eq:friedmann}
\end{equation}
In all the above equations, a dot means derivative with respect to
the cosmic time $t$, $a(t)$ is the scale factor, $H$ is the Hubble
parameter, $G$ is Newton's gravitational constant, and we use units in
which $c=1$.

\subsection{Dynamical system}
To study the dynamics of the system~(\ref{eq:eqs-motion}), it is
convenient to define new dimensionless variables as follows,
\begin{equation}
x \equiv \frac{\sqrt{4\pi G}}{\sqrt{3}H} \dot{\phi} \, , \quad y
\equiv \frac{\sqrt{4\pi G}}{\sqrt{3} H} m \phi\, , \quad z \equiv
\frac{m}{H} \, . \label{eq:new}
\end{equation}
The evolution equations~(\ref{eq:eqs-motion}) are then replaced by the
equations
\begin{subequations}
\label{eq:dynamical}
\begin{eqnarray}
x^\prime &=& - 3(1-x^2) x - z y \, , \label{eq:dynamical-a} \\
y^\prime &=& 3 x^2 y + z x \, , \label{eq:dynamical-b} \\
z^\prime &=& 3 x^2 z \, . \label{eq:dynamical-c}
\end{eqnarray}
\end{subequations}
where a prime denotes derivative with respect to the e-fold number $N
\equiv \ln a$, and we have used Eq.~(\ref{eq:friedmann}) in the form
$-\dot{H}/H^2 =3x^2$; on the other hand, the Friedmann constraint now reads
\begin{equation}
  x^2 +y^2 = 1 \, . \label{eq:friedmannxy}
\end{equation}
Eqs.~(\ref{eq:dynamical}) are written in the form of a dynamical
system, which helps to clarify the properties of the original
equations of motion.

Unlike the very well known case of an exponential
potential\cite{Copeland:1997et,Liddle:1998jc,Copeland:2006wr}, it is
not possible to take the Friedmann parameter $H$ out of the evolution
equations, and we are forced to take it into account through the
definition of the new variable $z$; notice that this new variable is,
by definition, a monotonic growing function
(see\cite{Boehmer:2008av,Matos:2008ag} for a similar situation in a
cosmological dynamical system).

The trajectories in the 3-dim phase space are located on the surface
of an infinite cylinder of unitary radius, and, strictly speaking, the
only critical point is the origin of coordinates. It is then evident
that the 3-dimensional system does not provide enough relevant
information about the dynamics of the original physical system.

However, one is usually interested in the evolution of variables $x$
and $y$, whose quadratic values represent the contributions of the
kinetic and potential energies, respectively, to the total energy
contents of the universe. Because of this, the Friedmann constraint
involves the values of $x$ and $y$ only, see Eq.~(\ref{eq:friedmann}),
and then all important physical features of the original system of
equations~(\ref{eq:eqs-motion}) are more tractable if we restrict
ourselves to the $xy$ plane of the phase space.

Thus, we are to study the reduced $2$-dim dynamical system provided by
Eqs.~(\ref{eq:dynamical-a}) and~(\ref{eq:dynamical-b}) only; and
variable $z$ will be considered a \emph{control parameter} all the
results presented below will depend upon. In what follows, and for
purposes of generality, we will study the unconstrained
system~(\ref{eq:dynamical-a}) and~(\ref{eq:dynamical-b}), and we shall
impose the Friedmann constraint~(\ref{eq:friedmannxy}) whenever is
necessary to clarify the nature of the critical points.

\subsection{Critical points}
The critical points of the $2$-dimensional system correspond to those
for which  $x^\prime = y^\prime =0$. The results can be summarized as
follows.

\emph{Null scalar field}. This point corresponds to the
origin of coordinates $x=y=0$. This point exists for any value of $z$;
but it should be stressed out that this point does not accomplish with
the Friedmann constraint~(\ref{eq:friedmannxy}), i.e., it is
inaccessible to the original physical system.

\emph{Scalar field domination}. There are four critical points with
coordinates given by
\begin{equation}
  x^2_0 = \frac{1}{2} \pm \frac{1}{2} \sqrt{1-\frac{4}{9} z^2} \, ,
  \quad y^2_0 = \frac{1}{2} \mp \frac{1}{2} \sqrt{1-\frac{4}{9} z^2}
  \, . \label{eq:atractor1}
\end{equation}
Notice that these critical points cannot be considered fixed in the
strict sense, as their values depend upon variable $z$; but it is
remarkable that they all lie, whenever their existance is allowed, on
the unitary circumference $x^2_0+y^2_0=1$.

For $z=0$, the four critical points are located at $\{\pm 1,0\}$ and
$\{0,\pm 1\}$; but as $z$ evolves to larger values, the points move
one to each other in the unitary circumference until they merge into
two points located at $\{\pm 1/\sqrt{2}, \mp 1/\sqrt{2} \}$ once
$z=3/2$. The critical points cease to exist for $z > 3/2$, see
also Fig.~\ref{fig:portrait}

\subsection{Stability analysis}
We now proceed to the stability analysis under the assumption that
variable $z$ is, as mentioned before, only a control
parameter. Following standard methods, we make small perturbations
around each one of the critical points of the form $\mathbf{x} =
\mathbf{x}_0 + \mathbf{u}$, where $\mathbf{x} = (x,y)$,  $\mathbf{u} =
(u_x,u_y)$ are the corresponding perturbations of the dynamical
variables, and a subscript '$0$' denotes the critical points.

The dynamical system in Eqs.~(\ref{eq:dynamical-a})
and~(\ref{eq:dynamical-b}) is of the form $\mathbf{x}^\prime =
\mathbf{f}(\mathbf{x},z)$, and then the evolution equations for the
perturbations are $\mathbf{u}^\prime = \mathcal{M} \mathbf{u}$, where
the stability matrix $\mathcal{M}$ has components
\begin{equation}
  \mathcal{M}_{ij} = \left. \frac{\partial f_i}{\partial x_j}
  \right|_{\mathbf{x}_0} = \left( 
  \begin{array}{cc}
    -3+9x^2_0 & -z \\
    6x_0 y_0 + z & 3x^2_0
  \end{array}
  \right) \, .
\end{equation}
We only need to calculate the eigenvalues of the stability matrix by
taking the values of $x_0$ and $y_0$ at each one of the critical
points; the results are summarized below.

\emph{Null scalar field}. The eigenvalues of the stability matrix in
this case are
\begin{equation}
  \omega_{1,2} = -\frac{3}{2} \pm \frac{3}{2} \sqrt{1-\frac{4}{9} z^2}
  \, .
\end{equation}
For any value $z > 0$, this critical point is stable; for values $0 <
z < 3/2$, the point is a node, whereas for values $3/2 < z$ the
point is a converging focus. Unfortunately, we have already mentioned
that this critical point does not accomplish the Friedmann
constraint~(\ref{eq:friedmannxy}), and then it is not of physical
interest.

\emph{Scalar field domination}. For the rest of the critical
points~(\ref{eq:atractor1}), the eigenvalues are given by
\begin{equation}
  \omega_1 = 6 x^2_0 \, , \quad \omega_2 = 3 \left( 2x^2_0 -1 \right)
  \, .
\end{equation}
It can be seen that all the critical points are unstable under small
perturbations whenever they exist, as the first eigenvalue is always
positive, $\omega_1 \geq 0$. However, the second eigenvalue is
negative (positive) for $x^2_0 < 1/2$ ($x^2_0 \geq 1/2$). This change
of sign does not make any difference for the stability of the critical
points, except in the case of trajectories subjected to the Friedmann
constraint~(\ref{eq:friedmannxy}), as we are to show now.

The general solution to the perturbations is of the form
\begin{equation}
  \mathbf{u} = C_1 \left(
  \begin{array}{c}
    x_0 \\
    y_0
  \end{array} \right) e^{\omega_1 t} + C_2 \left(
  \begin{array}{c}
    y_0 \\
    -x_0
  \end{array} \right) e^{\omega_2 t} \, ,
\end{equation}
where the terms in brackets are the normalized eigenvectors of the
stability matrix, and the $C$'s are arbitrary constants which are
determined from initial conditions. The Friedmann
constraint~(\ref{eq:friedmannxy}), to first order in perturbations,
reads
\begin{equation}
  x_0 u_x + y_0 u_y = C_1 e^{\omega_1 t} = 0 \, ,
\end{equation}
and then $C_1 =0$ at all times. In other words, the Friedmann
constraint does not allow the existence of radial perturbations.

Therefore, the nature of the critical points for any trajectory
constrained to the unitary circumference only depends on the value of
$\omega_2$. In consequence, there are two stable (attractor) points at
\begin{equation}
  x^2_s = \frac{1}{2} - \frac{1}{2} \sqrt{1-\frac{4}{9} z^2} \, ,
  \quad y^2_s = \frac{1}{2} + \frac{1}{2} \sqrt{1-\frac{4}{9} z^2} \,
  , \label{eq:attractor3}
\end{equation}
and there are two unstable points at
\begin{equation}
  x^2_u = \frac{1}{2} + \frac{1}{2} \sqrt{1-\frac{4}{9} z^2} \, ,
  \quad y^2_u = \frac{1}{2} - \frac{1}{2} \sqrt{1-\frac{4}{9} z^2} \,
  . \label{eq:attractor4}
\end{equation}

\begin{figure}[ht]
\centerline{\psfig{file=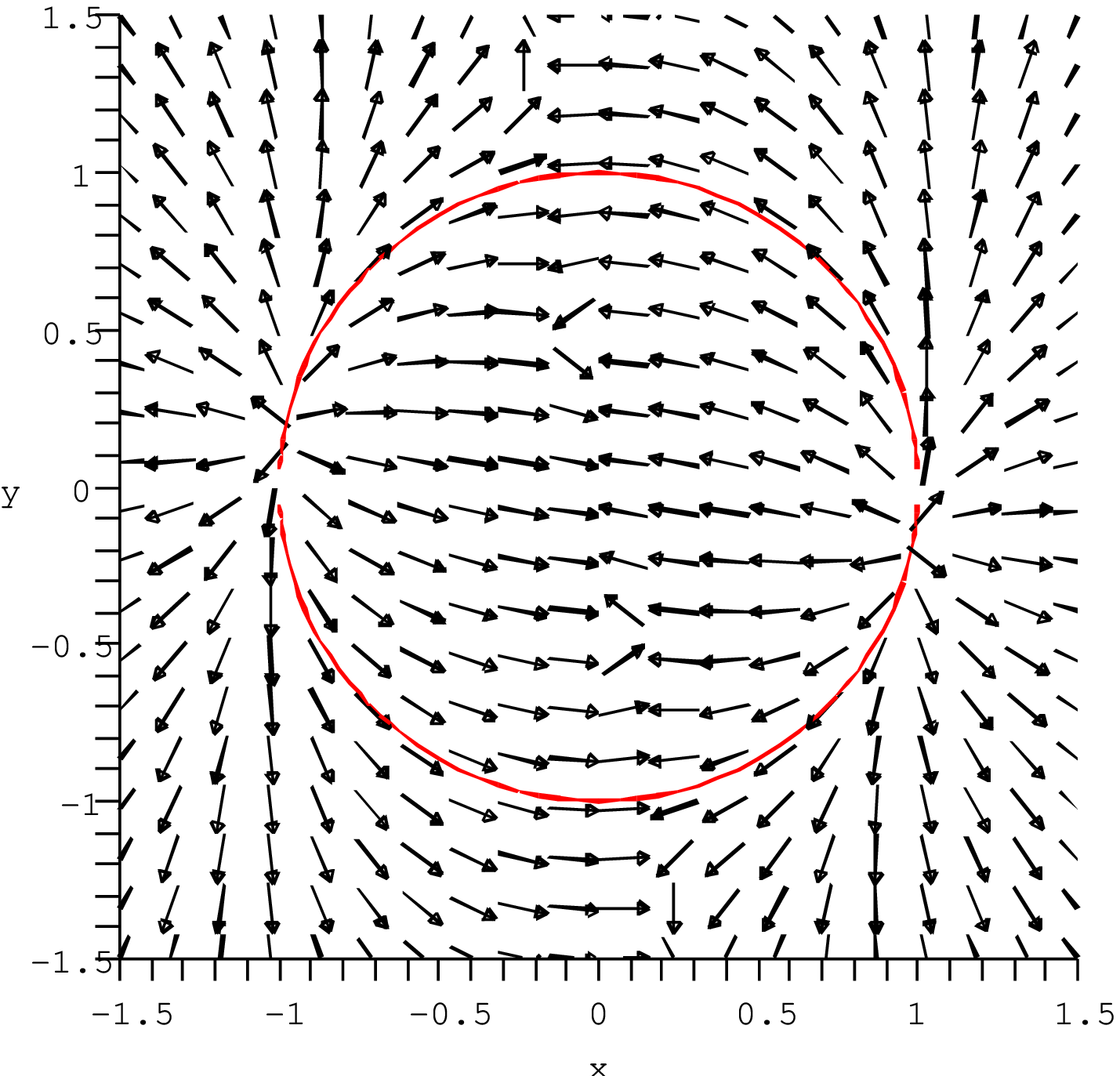,width=2.5in}
\psfig{file=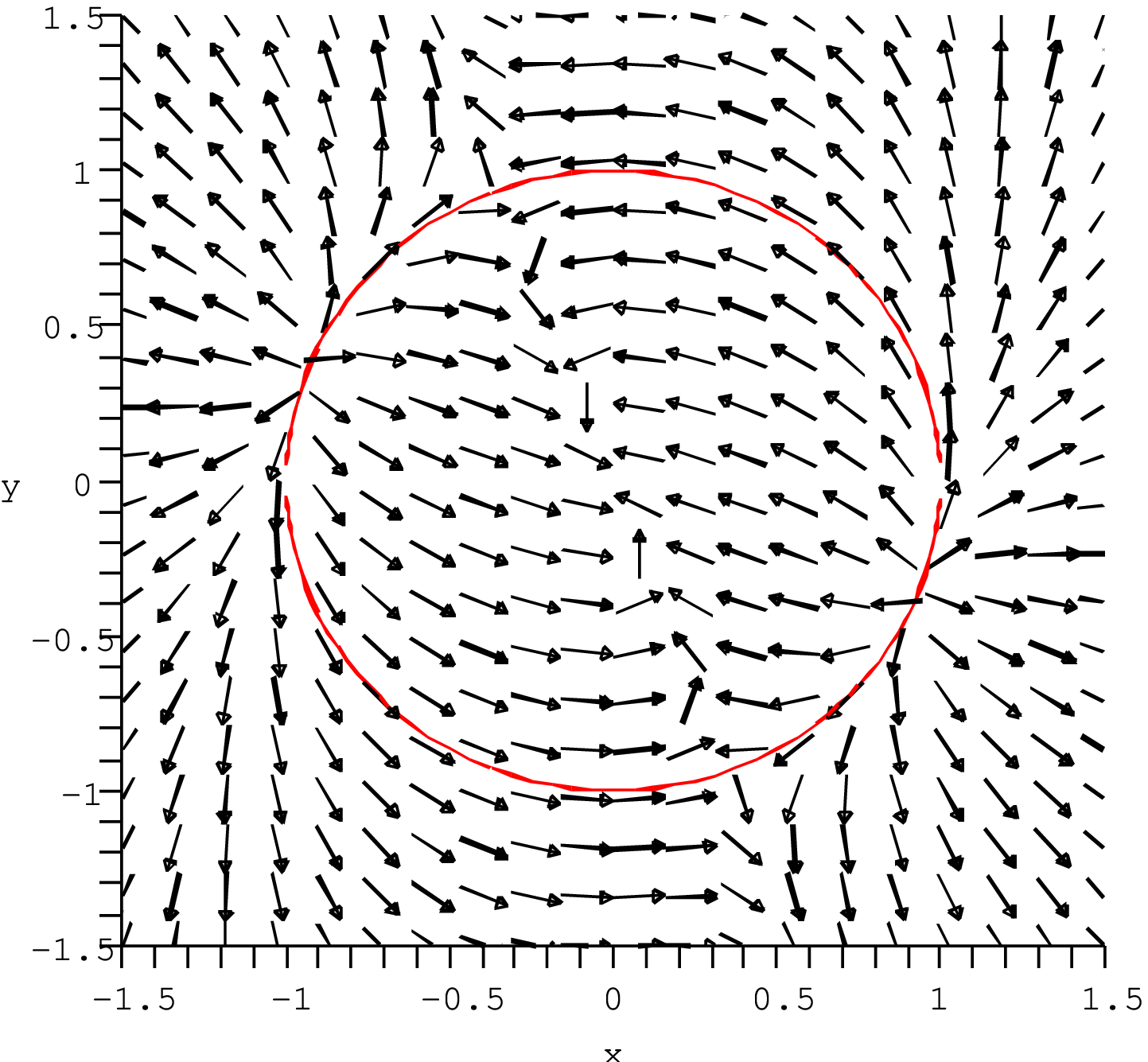,width=2.5in}}
\vspace*{3pt}
\centerline{(a)$z=0.5$ \hspace{5cm} (b)$z=1$}
\vspace*{3pt}
\centerline{\psfig{file=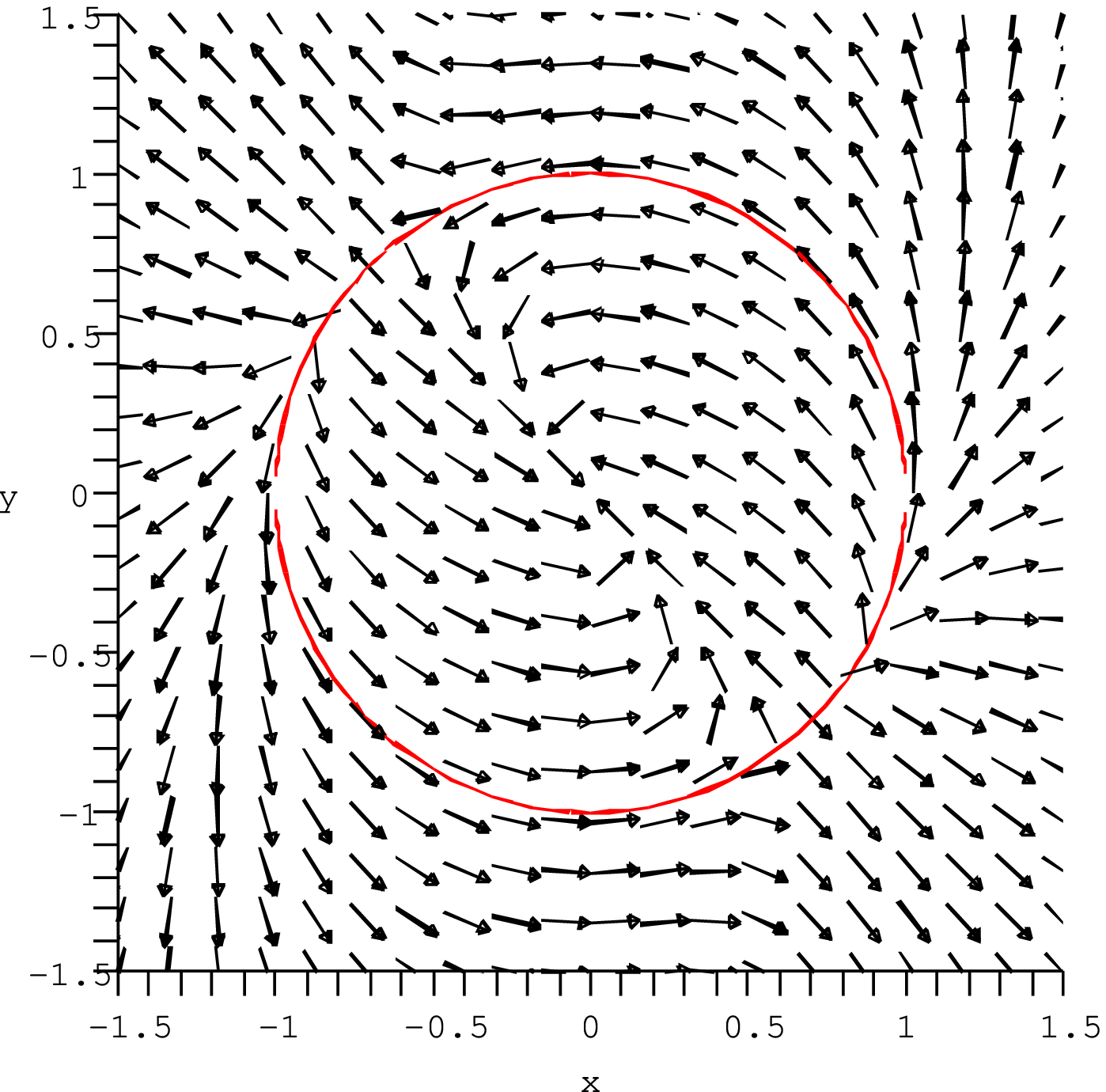,width=2.5in}
\psfig{file=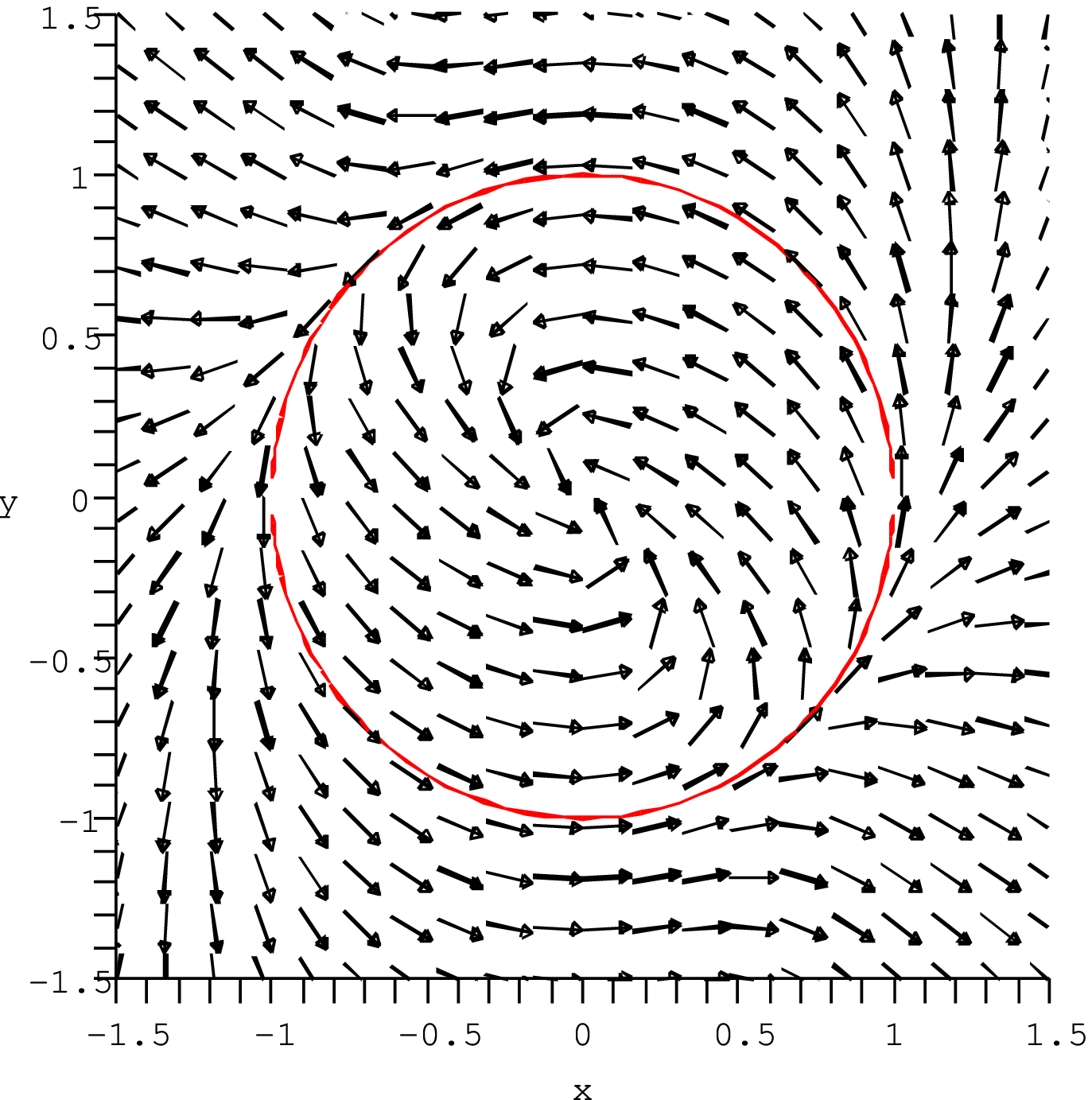,width=2.5in}}
\vspace*{3pt}
\centerline{(c)$z=1.5$ \hspace{5cm} (d)$z=2$}
\vspace*{3pt}
\caption{\label{fig:portrait} 2-dimensional phase portraits for the
  set of Eqs.~(\ref{eq:dynamical-a}) and~(\ref{eq:dynamical-b}), in
  which variable $z$ is taken as a free parameter. The arrows show the
  normalized velocity field with components $\mathbf{v}=(x^\prime,
  y^\prime)$, and the unitary circumpherence represents the Friedmann
  constraint~(\ref{eq:friedmannxy}). As expected, the origin of
  coordinates is a convergent focus in the four figures, and there are
  other four critical points given by Eq.~(\ref{eq:atractor1}), which
  are all unstable, located on the unitary circumference for $0< z <
  3/2$; the latter merge into two points at the special value
  $z=3/2$. See the text for more details.}
\end{figure}

Fig.~\ref{fig:portrait} show some sketches of the phase portrait, as
depicted in the mathematical package MAPLE, of our dynamical system on
the $xy$ plane for different values of $z$; the figures speak by
themselves and illustrate well the description of the critical points
made in the text above.

After the attractor points has disappeared, the trajectories move in
the counterclockwise direction with increasing angular frequency. To
leading order for large values of $z$, one can show that variables $x$
and $y$ obey a harmonic oscillator equation where variable $z$ plays
the role of the angular frequency. This fact makes the numerical
evolution very difficult to follow for times corresponding to $z >
2/3$.

\subsection{'Fixed point' (FP) approximation}
The method that was developed in the previous sections is going to be
called the 'fixed point' (FP) approximation. The reason is that the
2-dim system~(\ref{eq:dynamical-a}) and~(\ref{eq:dynamical-b}) is
not an autonomous one, and then the values $(x,y)$ for which the terms
on their r.h.s. vanish are not, in the strict sense, solutions of the
2-dim dynamical system.

However, the 2-dim velocity field vanishes at the critical points, see
Fig.~\ref{fig:portrait}, which indicates that they should have
attractor properties. In this respect, the standard perturbation
analysis gives correct information about the stability of the critical
points. It should be stressed out that the stability analysis is valid
only on constant-$z$ slices of the 3-dim phase space; in other words,
we have to neglect the time evolution of variable $z$. In consequence,
the FP is expected to work well only in the cases for which the value
of $z^\prime$, see Eq.~(\ref{eq:dynamical-c}), is small.

\begin{figure}[ht]
  \centerline{\psfig{file=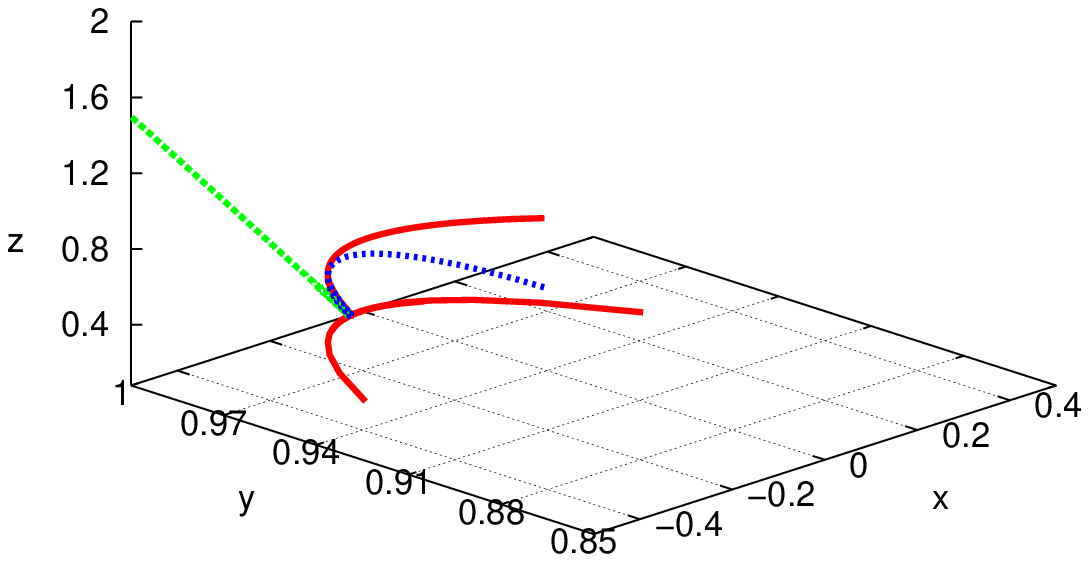,width=4.0in}}
  \vspace*{3pt}
  \centerline{\psfig{file=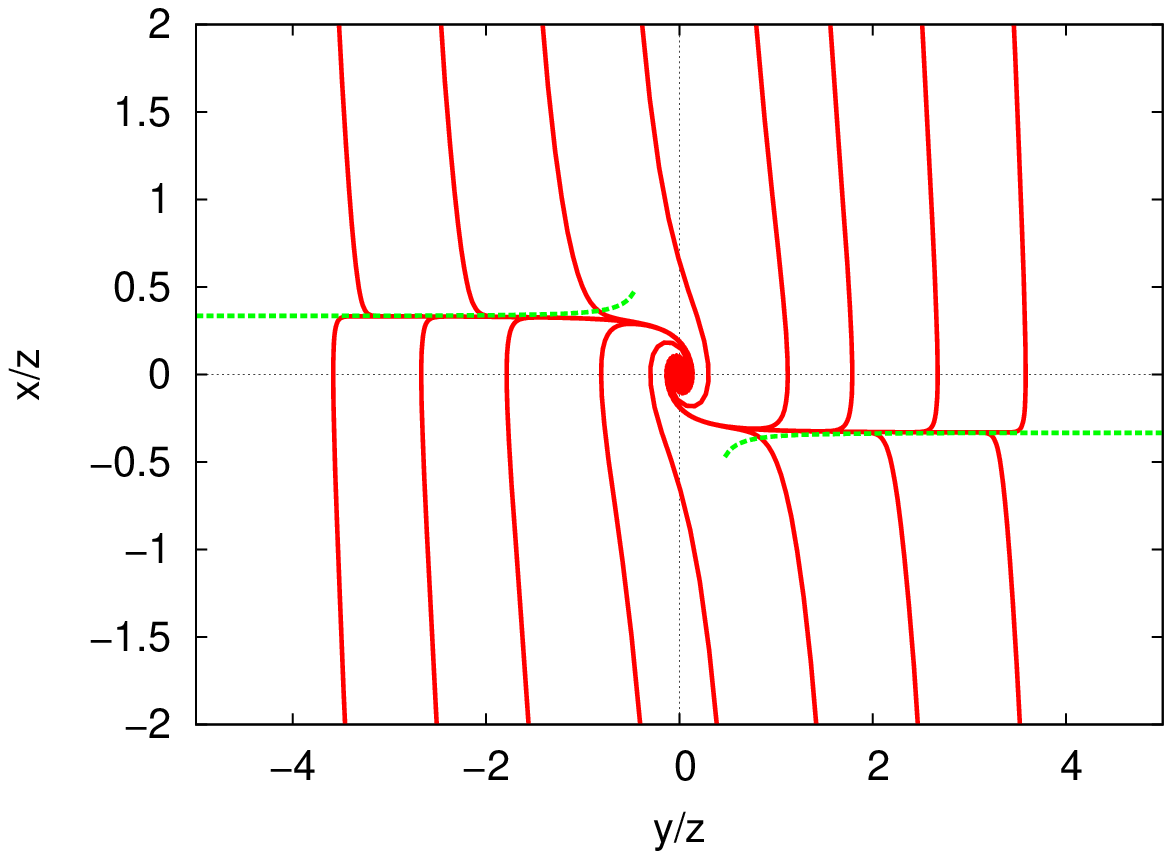,width=3.4in}}
  \vspace*{3pt}
  \caption{\label{fig:portrait2} (Top) Comparison between the 
    numerical solutions of the full 3-dim equations of
    motion~(\ref{eq:dynamical}) (red solid lines) and the fixed point (FP)
    (blue dotted line) and slow-roll (SR) (green dashed line)
    approximations. The numerical solutions were given initial
    conditions according to the Friedmann
    constraint~(\ref{eq:friedmannxy}) at $z_i=0.08$. The trajectories
    for the FP approximation were obtained from the attractor
    solutions~(\ref{eq:attractor3}), whereas those corresponding to
    the SR approximation are given by Eqs.~(\ref{eq:SR})
    in~\ref{sec:comparison-with-slow}. Initially, the numerical
    solution quickly reaches the attractor solution given by the FP
    approximation, but they differ at late times. However, the
    separation from the SR approximation is much larger;
    see~\ref{sec:comparison-with-slow} for more
    details. (Bottom) 2-dim phase space with the same
      variables used in Fig.~1 of Ref.~2. The red solid lines
      represent different solutions of the equations of
      motion~(\ref{eq:dynamical}), whereas the green dashed lines are
      the trajectories obtained from the stable critical
      points~(\ref{eq:attractor3}). Notice that the latter agree well
      with the inflationary separatrices described in Ref.~2. There is
      a small discrepancy close to the origin, which shows the
      break-down of the FP approximation at the late stages of
      inflation; see the text for more details.
  }
\end{figure}

An example is shown in Fig.~\ref{fig:portrait2} to illustrate that the
FP approximation is very good at early times, and that it differs from
the true evolution of the system at late times when $z^\prime$ cannot
be neglected. In any case, in inflationary studies one is interested
in the early evolution of a given model where the FP approximation is
particularly useful.

To finish this section, we would like to compare our method
  with that used in Ref.~\refcite{Belinsky:1985zd}, where inflationary
  solutions were shown to be characteristic of the equations of
  motion~(\ref{eq:eqs-motion}). Using other dynamical variables
  different to ours in Eqs.~(\ref{eq:new}), the inflationary
  solutions were identified with two horizontal separatrices on a
  2-dim phase space, see Fig.~1 in Ref.~\refcite{Belinsky:1985zd}. We 
  show in our Fig.~\ref{fig:portrait2} that those inflationary
  separatrices  are well described by our stable critical
  points~(\ref{eq:attractor3}). From the comparison above, we see that
  the FP approximation may provide more semi-analytical results than
  other methods.

\section{Inflationary dynamics}
 \label{sec:infl-dynam}

One of the issues cosmologists are most interested in are the
inflationary solutions provided by scalar field models. The
minimally-coupled scalar field endowed with a quadratic potential has
a long tradition as an inflationary model, and its predictions are
very well known, see for
instance\cite{Linde:1983gd,Liddle:2000cg,Bassett:2005xm,Alabidi:2006qa};
besides, it seems to be the best model to fit the available
observational data\cite{Komatsu:2008hk,Spergel:2006hy}, see
also\cite{Peiris:2003ff}. In this section, we will rewrite part of the
inflation formalism in terms of our new variables, and discuss the
role played by the critical points studied in the previous section.

\subsection{Attractor inflationary solutions}

The condition for an accelerated expansion is
\begin{equation}
  \frac{\ddot{a}}{a} = \dot{H} + H^2 > 0 \quad \iff \quad 3x^2 = -
  \frac{\dot{H}}{H^2} < 1 \, . \label{eq:acceleration}
\end{equation}
The description of the dynamics of a scalar field with a quadratic
potential we provided in Sec.~\ref{sec:math-backgr} confirms the
existence of attractor points that provides an accelerated expansion;
but that happens only as long as the attractor points accomplish with the
condition~(\ref{eq:acceleration}); that is, if $3 x^2_s < 1$.

As the position of the critical points depends upon $z$, we see that
the attractor points~(\ref{eq:attractor3}) become non-inflationary
for $z > z_{end} = \sqrt{2}$, where the subscript \emph{end} means the
time the attractor solution leaves the accelerated regime; the
corresponding values for the other variables are $x^2_{c,end} = 1/3$
and $y^2_{c,end} = 2/3$.

In fact, our calculations show that the scalar field value once the
attractor is no longer inflationary is 
\begin{equation}
  \phi_{end} = \sqrt{\frac{3}{4\pi}} \frac{y_{end}}{z_{end}} =
  \frac{m_{Pl}}{2\sqrt{\pi}} \, ,
\end{equation}
where $m_{Pl} \equiv G^{-1/2}$ is the Planck mass. This value
coincides with that provided by the slow-roll formalism of
inflation\cite{Belinsky:1985zd,Starobinsky78,Liddle:2000cg}.

\subsection{Number of e-folds during inflation}
Another point we investigated refers to the number of e-folds during
which inflation occurs, but for that we required the help of numerical
solutions. 

First of all, we have to set initial conditions for $\phi_i$,
$\dot{\phi}_i$ and $H_i$, which should be translated into the initial
conditions $x_i$, $y_i$ and $z_i$. The different physical constraints
to be imposed upon them are discussed in the following list.

\begin{itemize}
\item The first constraint to be taken into account is $\phi_i
  /m_{Pl} \sim \mathcal{O}(1)$. In order to obtain results as general
  as possible, we took the inequality
\begin{equation}
  \frac{\phi_i}{m_{Pl}} = \sqrt{\frac{3}{4\pi}} \frac{y_i}{z_i} < 10
  \, . \label{eq:const-1}
\end{equation}

\item The second constraint arises from the requirement that
  $\rho_{\phi,i} < m^4_{Pl}$, a constraint that is called the
  \emph{quantum boundary} in Ref.\cite{Belinsky:1985zd}. With the help
  of the Friedmann constraint~(\ref{eq:friedmann}), one finds that
  such constraint corresponds to the \emph{lower} bound $z_ i >
  m/m_{Pl}$, which is in turn equivalent to impose $H_i < m_{Pl}$.

\item The initial value $\dot{\phi}_i$ is found in terms of the
  Friedmann constraint $x^2_i = 1-y^2_i$, once $y_i$ is found from the
  above constraints.

\item Last constraint refers to inflationary solutions, and then $z_i
  < z_{end}$.
\end{itemize}

We noticed, in the diverse numerical experiments we performed, that
the number of inflationary $e$-folds mainly depends upon the initial
value $z_i$. This is because the dynamical system reaches the
attractor points very quickly, and then its subsequent evolution can
be well described by the motion of the attractor points themselves.

Thus, a formula for the number of inflationary $e$-folds is given by a
combination of Eqs.~(\ref{eq:dynamical-c}) and~(\ref{eq:attractor3}),
namely,
\begin{equation}
  N (z;z_i) \equiv \ln \left( \frac{a}{a_i} \right) \simeq
  \frac{2}{3} \int^z_{z_i} \frac{d \ln \tilde{z}}{\left( 1-\sqrt{1-
      4\tilde{z}^2/9} \right)} \, ,
  \label{eq:efolds}
\end{equation}
Needless to say, we have found very good agreement between our
numerical results and the analytic ones provided by
Eq.~(\ref{eq:efolds}).

The total number of e-folds in a given inflationary stage, defined
through $N_{total} \equiv N(z_{end};z_i)$ in Eq.~(\ref{eq:efolds}), is
uniquely determined by the initial value $z_i$, or equivalently, by
the initial value of the Hubble parameter $H_i$. If the total number
of $e$-folds should at least be $N_{total} > 60$, then the initial
condition should be $z_i < 0.15$; this guarantees that there is
sufficient inflation whatever the initial conditions on the scalar
field are, as long as they are in agreement with the Friedmann
constraint.

The attractor solution indicates that the end of the accelerating
stage happens once the Hubble parameter is $H_{end} = m/\sqrt{2}$, and
the oscillations begin a bit later, when $H_{osc} = 2m/3$, once the
attractor point disappears. These two events are almost instantaneous,
as the number of e-folds in between, as calculated again from
Eq.~(\ref{eq:efolds}), is $\Delta N=0.05$.

\subsection{Inflationary quantities}
We give here some instances about how we can write different
inflationary quantities of physical interest in terms of the attractor
values of our dynamical variables.

The first one is the amplitude of primordial quantum perturbations of
the inflaton field given by
\begin{equation}
  \delta^2_H (k) \equiv \frac{4}{25} \left( \frac{H}{\dot{\phi}}
  \right)^2 \left( \frac{H}{2\pi} \right)^2 = \frac{4}{75\pi}
  \frac{m^2}{m^2_{Pl}} \frac{1}{x^2_s(z) z^2} \, ,
  \label{eq:amplitude}
\end{equation}
where it is implicitly assumed, as usual, that all quantities on the
r.h.s. are evaluated at the time the corresponding wave number $k$
leaves the horizon, i.e. $k=aH$. Also, notice that the last term on
the r.h.s. should be calculated using the attractor solutions in
Eq.~(\ref{eq:attractor3}). 

The last statement is not trivial, as the standard procedure is to
calculate the values of $H$ and $\dot{\phi}$ in the slow-roll
regime. This is not our case, as we are giving the aforementioned
dynamical variables their values at the attractor points, and the
latter are not subjected to any order approximation. The same remark
applies to other quantities shown below.

The wave number $k$ in Eq.~(\ref{eq:amplitude}) can be given in terms
of the smallest scale that leaves the horizon at the end of inflation,
which we call $k_{end}$. In terms of our variables, it is easy to show
that
\begin{equation}
  \ln (k/k_{end}) = N(z;z_i) - N_{total} + \ln (z_{end}/z) \,
  . \label{eq:wavek}
\end{equation}
That given, we can calculate the spectral index $n(k)$ with the help
of Eqs.~(\ref{eq:amplitude}) and~(\ref{eq:wavek}), and then we obtain
\begin{equation}
  n(k) - 1 \equiv \frac{d\ln d^2_H(k)}{d\ln k} = - \frac{2\left[z^2-3x^2_s
  (z)\right]}{\left[1-3x^2_s(z)\right]\left[1-2x^2_s(z)\right]} \,
  , \label{eq:spectral}
\end{equation}
where, as before, the last term on the r.h.s. is the final expression
found from the attractor solution in terms of our dynamical
variables.

As said before, our calculations rely on exact inflationary attractor
solutions, so that we can proceed further and calculate higher order
  derivatives of the scalar power spectrum. For instance, the running
  of the spectral index $\alpha$ can be obtained just by one more
  parametric derivation of Eq.~(\ref{eq:spectral}),
\begin{equation}
  \alpha \equiv \frac{d n_s}{d\ln k} = \frac{4x^2_s(z) z^2\left[z^2 +
      9x^2_s(z) - 6\right]}{3\left[1 - 3x^2_s(z)\right]^3 \left[1 -
      2x^2_s(z)\right]^3} \, . \label{eq:run}
\end{equation}
As in other results shown before, this formula is evaluated at the
attractor solution. By contrast, the slow-roll formalism needs of a
second order calculation in the slow-roll parameters to get a precise
result\cite{Liddle:2000cg,Bassett:2005xm}.

For completeness, we give below the values of the aforementioned
inflationary quantities at a time corresponding to $60$ $e$-folds
before the end of inflation\cite{Liddle:2003as}. In terms of our
control parameter, that time corresponds to $z \simeq 0.15$, as
$N(\sqrt{2};0.15) = 60$, see Eq.~(\ref{eq:efolds}). The resulting
values are
\begin{subequations}
\begin{eqnarray}
  {\delta^2_H}^{(60)} & \simeq & 3.0105 \times 10^2 \frac{m^2}{m^2_{Pl}}
  \, , \\
  n^{(60)} & \simeq & 0.969659 \, , \\
  \alpha^{(60)} &=& - 4.65 \times 10^{-4} \, .
\end{eqnarray}
\end{subequations}
The above values can be compared to those inferred from observations
of the Cosmic Microwave Background. For example, from the measured
amplitude $\delta_H \simeq 1.9 \times 10^{-5}$\footnote{See
  Ref.\cite{Liddle:2006ev} and table of parameters in
  http://lambda.gscf.nasa.gov.} one obtains the known result for the
scalar field mass, which is $m \simeq 10^{-6}
m_{Pl}$\cite{Komatsu:2008hk,Spergel:2006hy,Alabidi:2006qa}.

\section{Conclusions \label{sec:conclusions}}
We have shown, with the help of the theory of dynamical systems, the
existence of attractor solutions in a scalar field theory minimally
coupled to gravity and endowed with a quadratic potential, the
simplest and most known of the monomial potentials in inflationary
theory. 

The attractor solutions and other subsequent results were given
explicitly in terms of the auxiliar variable $z= m/H$, which acted as
a free parameter and controlled the behavior of the critical
points. We called this method the \emph{fixed point} (FP)
approximation. In particular, we applied the formalism to the study of
inflationary dynamics with a quadratic potential. Our results are
explicitly based on attractor solutions, and then we were able to
calculate diverse inflationary quantities without the need of order
approximations.

Also, we would like to stress out that the conditions for an
inflationary stage are largely determined by the value of our control
variable $z$, as it alone determines the existence of an inflationary
attractor solution. This confirms the results of previous
  studies about the feasibility of inflationary solutions from
  arbitrary initial conditions\cite{Belinsky:1985zd}, see also the
  discussion in Refs.~\refcite{Hollands:2002yb,Kofman:2002cj}.

We believe that the FP approximation presented in this paper can be
used further to fully exploit the predictions of the simplest chaotic
inflationary model, and to have a more precise comparison with
cosmological
observations\cite{Cortes:2007ak,Pahud:2007gi,Alabidi:2006qa,Komatsu:2008hk}. The
FP approximation can be easily generalized to other inflationary
models\footnote{During the revision process of this work, we
  discovered that the FP approximation was successfully applied to a
  quartic monomial potential in Ref.~\refcite{Kiselev:2008zm}, where
  the authors also make a detailed comparison with the results
  obtained from slow-roll.}; this is work in progress that will be
published elsewhere.

\section*{Acknowledgments}
We thank Tonatiuh Matos and Gabriela Caldera-Cabral for useful
conversations. MR-I acknowledges partial support from Conacyt. This
work was partially supported by grants from Conacyt
(46195,47641,56946), DINPO, and PROMEP-UGTO-CA-3. This work is part of
the Instituto Avanzado de Cosmolog\'ia (IAC, Mexico) collaboration. 

\appendix

\section{Comparison with slow-roll}
\label{sec:comparison-with-slow}

In this appendix, we are going to make the comparison of the 'fixed
point' (FP) aproximation to the slow-roll (SR) one. The SR
approximation for the equations of motion of a scalar field in a FRW
metric and endowed with a quadratic potential
is\cite{Liddle:2000cg,Bassett:2005xm} (see also
  Refs.~\refcite{Belinsky:1985zd,Starobinsky78} for an earlier study
  of the SR equations)
\begin{equation}
    \dot{\phi} \simeq - \frac{m^2 \phi}{3H} \, , \quad
    H^2 \simeq \frac{8\pi G}{6}m^2 \phi^2 \, .
\end{equation}
In terms of our auxiliary variables~(\ref{eq:dynamical}), equivalent
expressions for the SR equations are
\begin{equation}
  x_{SR} \simeq - \frac{z}{3} \, , \quad y^2_{SR} \simeq 1 \,
  . \label{eq:SR}
\end{equation}

From Eqs.~(\ref{eq:attractor3}), it can be seen that the SR solution
is a first order approximation to the FP one. Actually, in the limit
$z \ll 1$ we get
\begin{equation}
  x_s \simeq \mp \frac{z}{3} \, , \quad y_s \simeq \pm \left( 1 -
  \frac{z^2}{18} \right) \, . \label{eq:sr-limit}
\end{equation}
The second equation shows that the correction to the usual SR
assumption $y_{SR} \simeq 1$ is of the second order. Likewise, the
values of the inflationary quantities in the limit $z \ll 1$ are, from
Eqs.~(\ref{eq:amplitude}),~(\ref{eq:spectral}), and~(\ref{eq:run}),
\begin{equation}
  d^2_H \simeq \frac{36}{75 \pi} \frac{m^2}{m^2_{\rm Pl}} z^{-4} \, ,
  \quad n(k) -1 \simeq - \frac{4}{3} z^2 \, , \quad \alpha \simeq
  -\frac{8}{9} z^4 \, .
\end{equation}
The above formulas can be written in terms of the number of $e$-folds
given by Eq.~(\ref{eq:efolds}); to first order we find
$N(\sqrt{2},z_i) \simeq (3/2) z^{-2}_i$ in the limit $z_i \ll
1$. If we use this last result in Eqs.~(\ref{eq:sr-limit}), we find
the usual expressions of the SR approximation for the different
inflationary quantities in terms of the number of
$e$-folds\cite{Liddle:2000cg}; see also\cite{Kiselev:2008zm}.

Some remarks are in turn. First one is that the Friedmann constraint
is not fully accomplished in the SR approximation, as we have seen
that the deviations are of second order. Second, it is manifest that
SR equations~(\ref{eq:SR}) are not solutions of the original equations of
motion. Moreover, the SR points are not even critical points of the
dynamical system~(\ref{eq:dynamical}), and so it is difficult to study
their attractor properties using standard techniques of dynamical
systems.

The SR and the FP trajectories coincide in the very early universe, in
the limit $z \ll 1$. But even for moderate small values of $z$, the
deviations of SR from FP start to be significant, see
Fig.~\ref{fig:portrait2}, and then one has to include correction terms
in the SR approximation at different
levels\cite{Liddle:2000cg,Bassett:2005xm}.



\end{document}